 \documentclass[twocolumn]{aastex631}
\usepackage{amsmath}
\usepackage{breqn}

\font\manual=manfnt at 7pt \def\dbend{\hbox{\raise0.9ex\hbox{\manual\char127\hspace{0.6em}}}}

\providecommand{\e}[1]{\ensuremath{\times 10^{#1}}}

\newcounter{INTERNALionstage}

\def\gtsim{\mathrel{\hbox{\rlap{\hbox{\lower4pt\hbox{$\sim$}}}\hbox{$>$}}}}
\def\lesssim{\mathrel{\hbox{\rlap{\hbox{\lower4pt\hbox{$\sim$}}}\hbox{$<$}}}}

\def\A{{\rm\thinspace \AA}}

\def\g{{\rm\thinspace g}}

\def\K{{\rm\thinspace K}}

\def\m{{\rm\thinspace m}}

\def\s{{\rm\thinspace s}}

\def\W{{\rm\thinspace W}}

\def\h0{\mbox{{\rm H}$^0$}}
\DeclareMathAlphabet{\vib}{OML}{cmm}{m}{it}
\newcommand*{\satellite}[1]{\textit{#1}}
\newcommand*{\xmm}{\satellite{XMM-Newton}}
\newcommand*{\chandra}{\satellite{Chandra}}
\newcommand*{\suzaku}{\satellite{Suzaku}}

\renewcommand{\thefootnote}{\fnsymbol{footnote}}

\shorttitle{$\chandra$ observations of A98}
\graphicspath{{./}{figures/}}

\begin{document}

\title{Discovery of a pre-merger shock in an intercluster filament in Abell 98}

\author[0000-0002-5222-1337]{Arnab Sarkar}
\affiliation{University of Kentucky, 505 Rose street, Lexington, KY 40506, USA}
\affiliation{Center for Astrophysics $\vert$ Harvard \& Smithsonian, Cambridge, MA 02138, USA}
\email{arnab.sarkar@uky.edu, arnab.sarkar@cfa.harvard.edu}

\author[0000-0002-3984-4337]{Scott Randall}
\affiliation{Center for Astrophysics $\vert$ Harvard \& Smithsonian, Cambridge, MA 02138, USA}

\author{Yuanyuan Su}
\affiliation{University of Kentucky, 505 Rose street, Lexington, KY 40506, USA}

\author[0000-0001-9266-6974]{Gabriella E. Alvarez}
\affiliation{Harvard-Smithsonian Center for Astrophysics, 60 Garden Street, Cambridge, MA 02138, USA}

\author{Craig Sarazin}
\affiliation{University of Virginia, Charlottesville, VA 22904, USA}

\author{Paul Nulsen}
\affiliation{Center for Astrophysics $\vert$ Harvard \& Smithsonian, Cambridge, MA 02138, USA}
\affiliation{ICRAR, University of Western Australia, 35 Stirling Hwy, Crawley, WA 6009, Australia}

\author{Elizabeth Blanton}
\affiliation{Boston University, Boston, MA 02215, USA}

\author{William Forman}
\affiliation{Center for Astrophysics $\vert$ Harvard \& Smithsonian, Cambridge, MA 02138, USA}

\author{Christine Jones}
\affiliation{Center for Astrophysics $\vert$ Harvard \& Smithsonian, Cambridge, MA 02138, USA}

\author{Esra Bulbul}
\affiliation{Max Planck Institute for Extraterrestrial Physics, Giessenbachstrasse 1, 85748 Garching, Germany}

\author{John Zuhone}
\affiliation{Center for Astrophysics $\vert$ Harvard \& Smithsonian, Cambridge, MA 02138, USA}

\author{Felipe Andrade-Santos}
\affiliation{Center for Astrophysics $\vert$ Harvard \& Smithsonian, Cambridge, MA 02138, USA}

\author{Ryan E. Johnson}
\affiliation{Gettysburg College, Gettysburg, PA 17325, USA}

\author{Priyanka Chakraborty}
\affiliation{Center for Astrophysics $\vert$ Harvard \& Smithsonian, Cambridge, MA 02138, USA}



\begin{abstract}
We report the first unambiguous detection 
of an axial merger shock in the early-stage 
merging cluster Abell~98
using deep (227 ks) $\chandra$ observations. 
The shock is about 420 kpc south from the 
northern subcluster of Abell 98, 
in between the northern and central subclusters, 
with a 
Mach number of 
$\cal{M}$ $\approx$ 2.3 $\pm$ 0.3.
Our discovery of the axial merger shock
front unveils a critical epoch in 
the formation of a massive galaxy cluster, when 
two subclusters are caught 
in the early phase of the merging process.
We find that the electron temperature in the post-shock 
region favours the instant collisionless model,
where electrons are strongly heated 
at the shock front, by interactions with the magnetic field.
We also report on the detection of an 
intercluster gas filament, with a temperature of
$kT$ = 1.07 $\pm$ 0.29 keV, along the merger axis of Abell~98.
The measured properties of the gas in the filament are consistent
with previous observations and numerical simulations of the
hottest, densest parts of the warm-hot intergalactic medium (WHIM),
where WHIM filaments interface with the virialization regions of galaxy clusters.


\end{abstract}

\keywords{Galaxy cluster --- Shock front --- Filament --- Cosmology}

\section{Introduction} \label{sec:intro}
Clusters of galaxies 
are primarily assembled and grow
via accretion, gravitational infall, and mergers of 
smaller substructures
and groups. In such merging events,
a significant fraction of kinetic energy
dissipates (on a Gyr timescale)
in the
intra-cluster medium (ICM) via 
shocks and turbulence \citep{1999ApJ...521..526M}.
Such shocks are the major heating
sources for the X-ray emitting 
ICM plasma
\citep{2007PhR...443....1M}.
Shock fronts provide an essential
observational tool in probing
the physics of transport processes
in the ICM, including
electron-ion equilibration and thermal
conduction, magnetic fields, and turbulence
\citep[e.g.,][]{2006ESASP.604..723M,2007PhR...443....1M,2016MNRAS.463.1534B,2018MNRAS.476.5591B}.

Despite the intrinsic interest and 
significance of merger shocks, 
X-ray observations of merger shocks with a sharp density 
edge and an unambiguous jump in temperature are relatively rare.  
Currently, only a 
handful of merger shock fronts have been identified by $\chandra$, 
such as 1E 0657-56 \citep{2002ApJ...567L..27M}, 
Abell 520 \citep{2005ApJ...627..733M},
Abell 2146 \citep{2010MNRAS.406.1721R,2012MNRAS.423..236R},
Abell 2744 \citep{2011ApJ...728...27O},
Abell 754 \citep{2011ApJ...728...82M},
Abell 2034 \citep{2014ApJ...780..163O},
Abell 665 \citep{2016ApJ...820L..20D}.

Cosmic filaments are thought 
to connect the large-scale structures of our universe
\citep[e.g.,][]{2006MNRAS.370..656D,2008A&A...482L..29W,2018ApJ...858...44A,2021A&A...647A...2R}. 
Several independent searches for baryonic mass 
have confirmed
discrepancies in baryonic 
content between the high and low redshift universe
\citep[e.g.,][]{1998ApJ...503..518F,2004ApJ...616..643F},
with fewer baryons being detected in the local
Universe.
They concluded that a significant fraction of these 
missing baryons may be ``hidden''
in the WHIM, 
in cosmic filaments that connect clusters and groups. 
The WHIM has a temperature in the 10$^5$-10$^7$ K 
(or, kT$\sim$0.01-1 keV) regime, and relatively 
low surface brightness
\citep[e.g.,][]{1999ApJ...514....1C,1999ApJ...511..521D,2001ApJ...552..473D,2011ApJ...731....6S}.

Abell 98 (hereafter A98) is a richness class III
early-stage merger with three
subclusters:
central (A98S; $z\approx$ 0.1063), 
northern (A98N; $z\approx$ 0.1042), 
and southern (A98SS; $z\approx$ 0.1218).  
The northern and southern subclusters
are at projected distances of 
$\sim$ 1.1 Mpc 
and 1.4 Mpc from the central subcluster,
respectively
\citep[e.g.,][]{1989ApJS...70....1A,1999ApJ...511...65J,1994ApJ...423...94B,2014ApJ...791..104P}.
The central subcluster is undergoing a separate 
late-stage merger, with two distinct X-ray cores.
Previous observations using $\chandra$ and $\xmm$ 
showed that A98N is a cool core cluster with 
a marginal detection of a warm gas arc, 
consistent with the presence of a 
leading bow shock, but that the exposure 
time was insufficient to confirm this feature
\citep{2014ApJ...791..104P}

To investigate further, we analyzed
deep ($\sim$227 ks) $\chandra$ observations
of A98N and A98S. 
In this letter, 
we report on the detection of an intercluster 
filament connecting A98N and A98S, 
and of a leading bow shock in the 
region of the filament, 
associated with the early-stage merger 
between A98N and A98S.
We adopted a cosmology of
$H_0$ = 70 km s$^{-1}$ Mpc$^{-1}$, 
$\Omega_{\Lambda}$ = 0.7, 
and $\Omega_{\rm m}$ = 0.3,
which gives a scale of 1$''$ = 1.913 kpc
at the redshift $z$ = 0.1042 of A98.
Unless otherwise stated, all reported error bars 
are at 90\% confidence level.\\

\section{Data analysis} \label{sec:data}
Abell 98 was observed 
twice with $\chandra$
for a total exposure time of $\sim$ 227 ks. 
The observation logs are listed 
in Table \ref{tab:obs_log}.
We discuss the detailed data reduction 
processes in Section \ref{sec:supp_data}.

\subsection{Imaging analysis} \label{sec:imaging}
\begin{figure*}[h!]
\centering
    \includegraphics[width=0.6\textwidth]{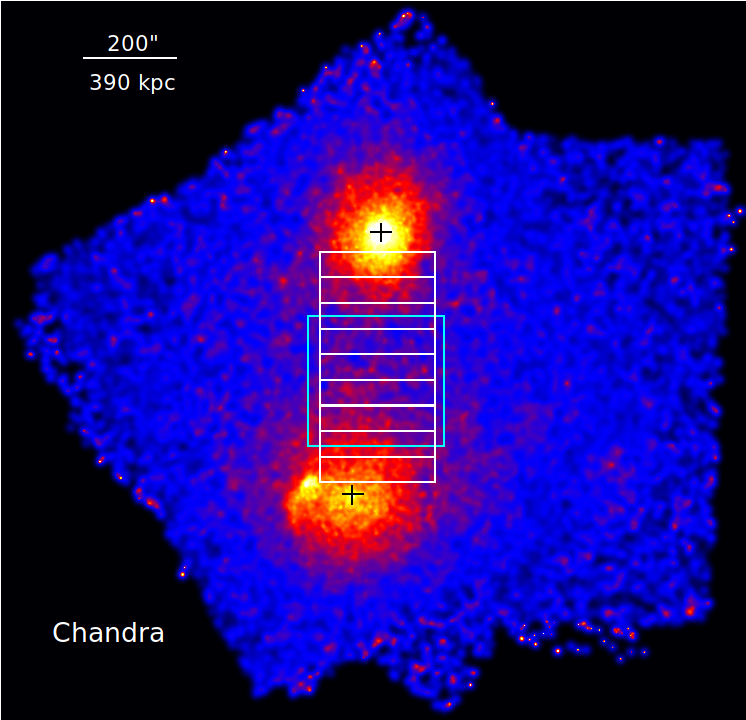}
    \caption{{\it Chandra}
exposure corrected and background subtracted
image of A98 in the 0.5--2 keV
energy 
band.
White and cyan 
regions are used for the analysis 
of the filament emission.
Centers of A98N and A98S are marked
in black.
\label{fig:bridge}}
\end{figure*}

\begin{figure*}
\gridline{\fig{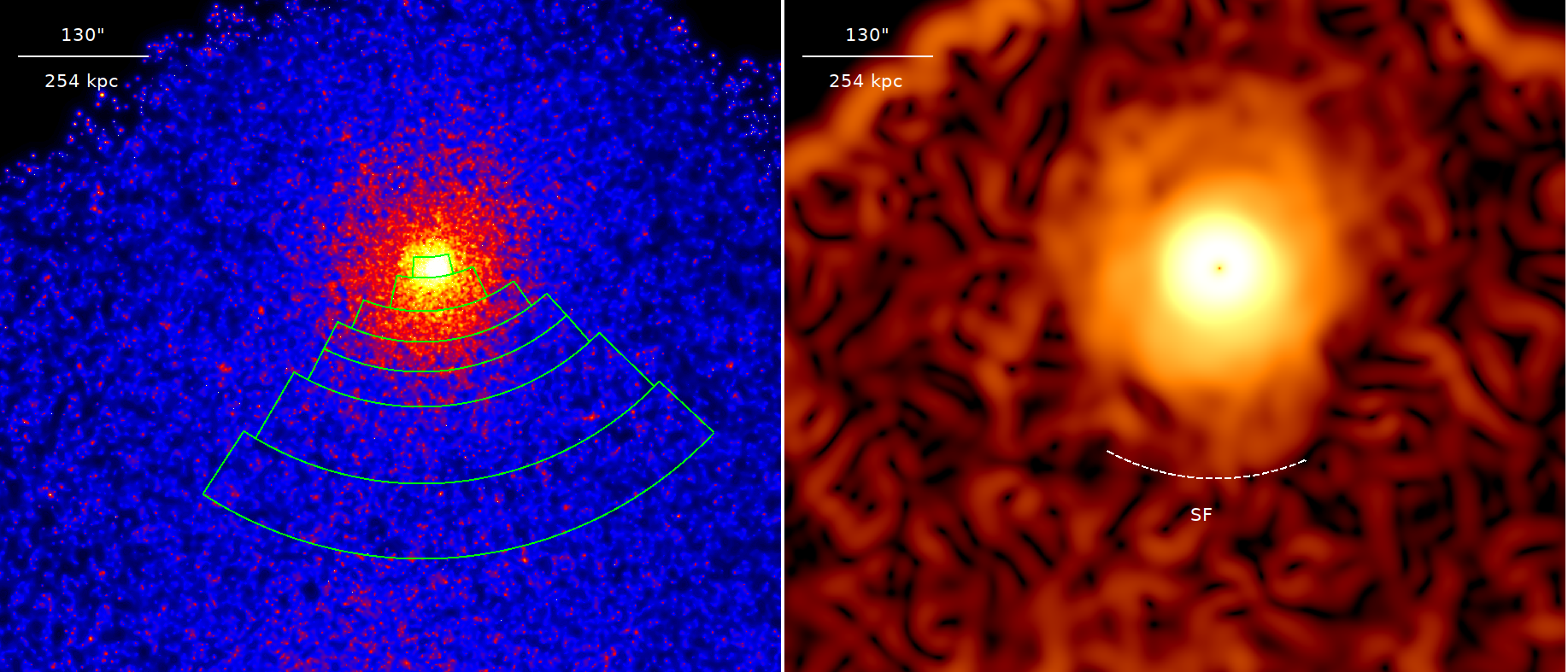}{1.0\textwidth}{}
        }
\vspace{-25pt}
\gridline{\fig{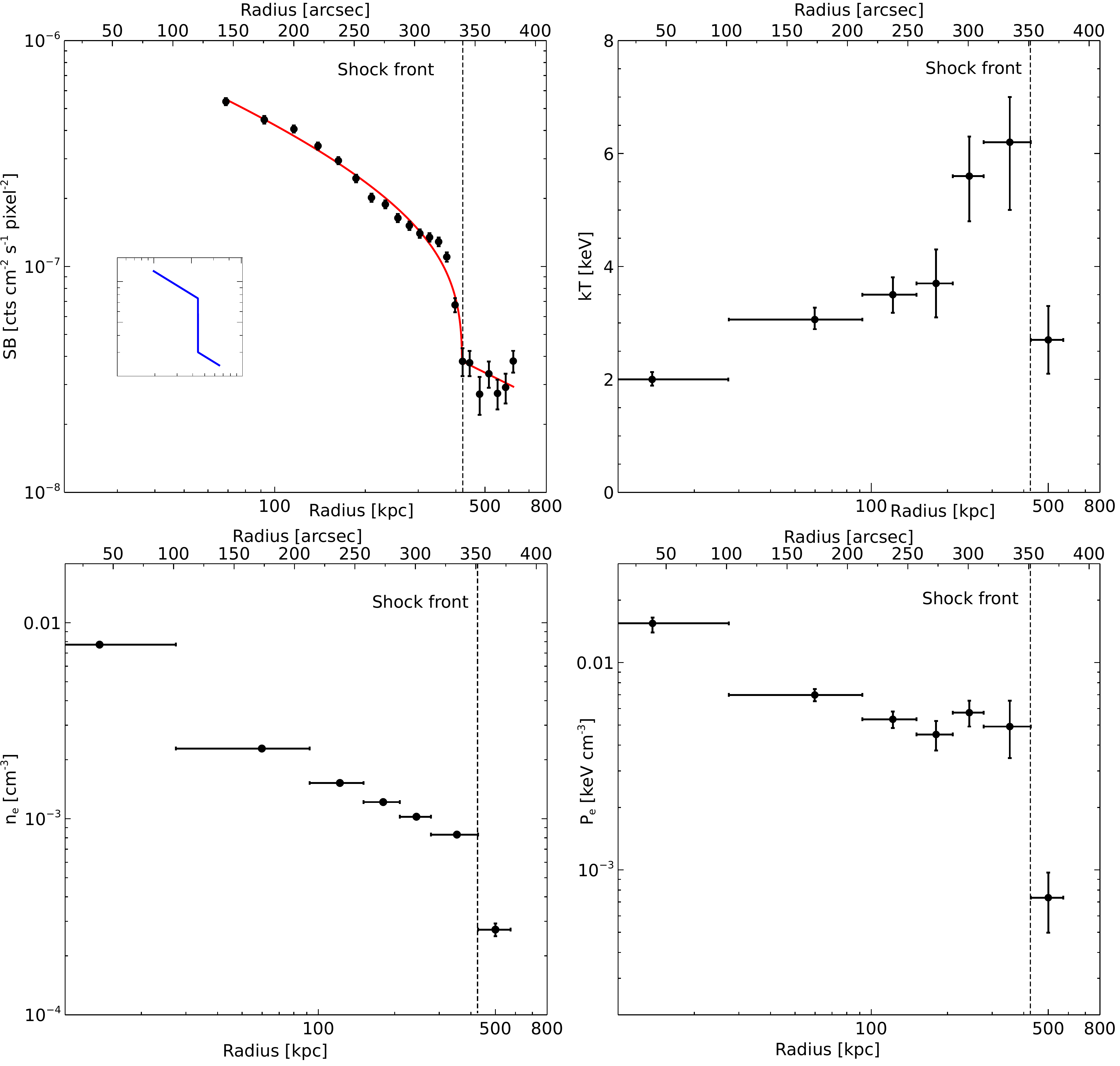}{0.8\textwidth}{}
        }
\caption{{\it Top-left}: 
Similar to Figure \ref{fig:bridge} but
zoom in to A98N and smoothed with $\sigma$=2$\arcsec$ Gaussian.
Green regions used for spectral analysis.
{\it Top-right}: 
GGM image of A98N
with $\sigma$=16$\arcsec$ Gaussian kernel.
White dashed curve shows southern shock front.
{\it Middle left}:
surface brightness profile of A98N 
in south direction with 1-$\sigma$ errorbars. 
Inset figure shows the 3D density profile.
{\it Middle-right}: 
projected temperature,
{\it bottom-left}: 3D density,
{\it bottom-right}: pressure profiles of A98N
in south direction.
\label{fig:image_edge_fitting}}
\end{figure*}

The image showing the both A98N and A98S 
in the 0.5-2 keV energy band
is shown
in Figure \ref{fig:bridge}.
We derived a
Gaussian Gradient Magnitude (GGM) 
filtered image of A98, zooming in the northern subcluster,
as shown in 
Figure \ref{fig:image_edge_fitting}. 
The GGM-filtered image provides a powerful tool to reveal 
sub-structures and any associated sharp features in the 
cluster core,
as well as at larger cluster radii \citep{2016MNRAS.461..684W}. 
The intensity of the GGM images indicates 
the slope of the local surface brightness 
gradient, with steeper gradients showing 
up as brighter regions.
The GGM image we present here is 
filtered on a length scale of 32 pixels. 
Each pixel is 
0.492$\arcsec$ wide. 
We observe 
a rapid change in the magnitude
of the surface brightness
gradient 
at $\sim$ 400 kpc to the south
of the A98N center,
as seen in Figure 
\ref{fig:image_edge_fitting}.

GGM images often show artifacts from the filtering 
process, particularly in
low surface brightness regions. 
Therefore, we next test for the 
presence of the edge feature indicated in
Figure \ref{fig:image_edge_fitting} 
by extracting a 
surface brightness profile 
across the edge. 
Figure \ref{fig:image_edge_fitting}
shows the resulting radial
surface brightness profile
as a function of distance
from the A98N core in the 0.5-2 keV energy band.
We observe an edge in the
X-ray surface brightness profile at 
about 420 kpc 
{(0:46:23.14, +20:33:46.32)} 
from the A98N core. 
The distance of the edge from the A98N core
is consistent with the
edge observed in the GGM image, which
suggests the abrupt change in the gradient
corresponds to the surface brightness
edge.

The shape of the extracted surface brightness profile 
across the edge 
is consistent with what is expected from a projection of a 
3D
density discontinuity \citep{2000ApJ...541..542M}. 
To quantify the 
surface brightness edge,
we fit the surface brightness 
profile 
by projecting a 3D discontinuous double power-law
model along the line of sight, defined as 
\begin{equation}
    n_e(r) \propto
    \begin{cases}
    \left(\frac{r}{r_{edge}}\right)^{-\alpha_{1}}, & \text{if $r < r_{edge}$}\\
    \frac{1}{jump} \left(\frac{r}{r_{edge}}\right)^{-\alpha_{2}}, & \text{if $r \geq r_{edge}$}
    \end{cases}
\end{equation}
where $n_e(r)$ is the 3D electron density at a radius $r$, 
$r_{edge}$
is the radius of the putative edge, $jump$ is the 
density jump factor, and $\alpha_1$ and $\alpha_2$
are the slopes before and after the edge, respectively. 
{A constant term 
was also added to the model to account
for any residual background, 
after blanksky subtraction.
The best-fit value of this
term was consistent with zero,
suggesting we successfully eliminated sky
and particle background.}
We project the estimated emission 
measure profile onto the sky plane and 
fit the observed surface 
brightness profile by 
using least square fitting
technique with 
$\alpha_1$, $\alpha_2$, $r_{edge}$, and the 
$jump$ as free parameters.
Figure 
\ref{fig:image_edge_fitting} 
shows the best-fit model and 
the 3D density profile (inset). 
The best-fit power-law indices
across the edge
are $\alpha_1$ = 0.76 $\pm$ 0.01
and $\alpha_2$ = 0.83 $\pm$ 0.02, 
respectively ($\chi^{2}$/dof = 57/21).
The density 
jumps, across the edge, by a factor 
of $\rho_2/\rho_1$ = 2.5 $\pm$ 0.3,  
where suffix 2 and 1 represent the regions
inside and outside of the front.
Assuming that the edge is a shock front, 
this density
jump corresponds to 
a Mach number of 
$\cal{M}$ = 2.3 $\pm$ 0.3,
estimated from the
Rankine-Hugoniot jump condition,
defined as
\begin{equation}
    {\cal{M}} = \left [\frac{2 r}{\gamma + 1 - r\left (\gamma - 1 \right)}\right]^{\frac{1}{2}},
\end{equation}
where $r$ = $\rho_2/\rho_1$ and for a monoatomic gas 
$\gamma$ = 5/3.
The edge radius, obtained from the fit, is 420 $\pm$ 25 kpc.
We estimated the uncertainties
by allowing 
all the other model parameters to vary freely.
The best-fit edge radius is consistent with the
distance of the GGM peak from the A98N center.

\subsection{Spectral analysis} \label{sec:spectral}
To measure the temperature across the surface 
brightness edge, the southern sector was divided into
seven regions, as shown in Figure
\ref{fig:image_edge_fitting}.
Each region contains a minimum 
of 2300 background-subtracted counts. 
We set this lower limit to guarantee adequate counts 
to measure the temperature uncertainty within 25\% 
in the faint region at a 90\% confidence level. 
For each region, we extracted spectra from individual 
observations and fitted them simultaneously. 
The spectra were grouped to contain a minimum of 20 counts 
per spectral channel. 
The blanksky-background spectra 
were subtracted from the 
source spectra before fitting \citep{2016ApJ...820L..20D}. 
We fitted the spectra from each region to an absorbed 
single-temperature thermal emission model, PHABS(APEC) 
\citep{2001ApJ...556L..91S}. 
The redshift was fixed to $z$=0.1042, and the absorption was 
fixed to the Galactic value of 
$N_{H}$=3.06$\times$10$^{20}$ cm$^{-2}$ 
\citep{2005A&A...440..775K}. 
The spectral fitting was performed using {\tt XSPEC} in the 0.6-7 keV 
energy band, 
and the best-fit parameters were obtained by 
reducing C-statistics. We fixed the metallicity 
to an average value of 0.4 Z$_{\odot}$ since 
it was poorly constrained if left free
\citep{2010MNRAS.406.1721R}.
We adopted solar abundance table of
\citet{2009ARA&A..47..481A}.

Figure \ref{fig:image_edge_fitting}
shows the best-fit projected temperatures. 
The projected temperature increases steadily from 
the A98N core up to $\sim$
200 kpc, then jumps to a peak of $\sim$ 6.4$^{+1.0}_{-1.5}$ keV at $\sim$ 420 kpc, 
which then plummets to 2.7$^{+0.5}_{-0.5}$ keV beyond the surface brightness edge.
Across the edge, the temperature decreases by a factor of 
$\sim$ 2.3 $\pm$ 0.6, 
confirming the outer edge as a shock front. 
{We estimated the electron
pressure by combining the 
temperature and electron density, 
as shown in Figure \ref{fig:image_edge_fitting}. 
As expected, we found a significant decrease in pressure 
by a factor of $\sim$ 7 $\pm$ 2 at the shock front.}
The observed temperature drop corresponds to a 
Mach number of $\cal{M}$ = 2.2 $\pm$ 0.4, 
consistent with the Mach number derived from the density 
and pressure jump. 
Mach numbers from both methods are consistent with
each other, bolstering the shock detection.
We estimated a pre-shock sound speed of $c_s$ $\sim$ 
848 $\pm$ 80 km/s, 
which gives a shock speed of 
$v_{shock}$ = ${ c_s} \cal{M}$
$\approx$ 1900 $\pm$ 180 km/s.

\begin{figure*}

\gridline{\fig{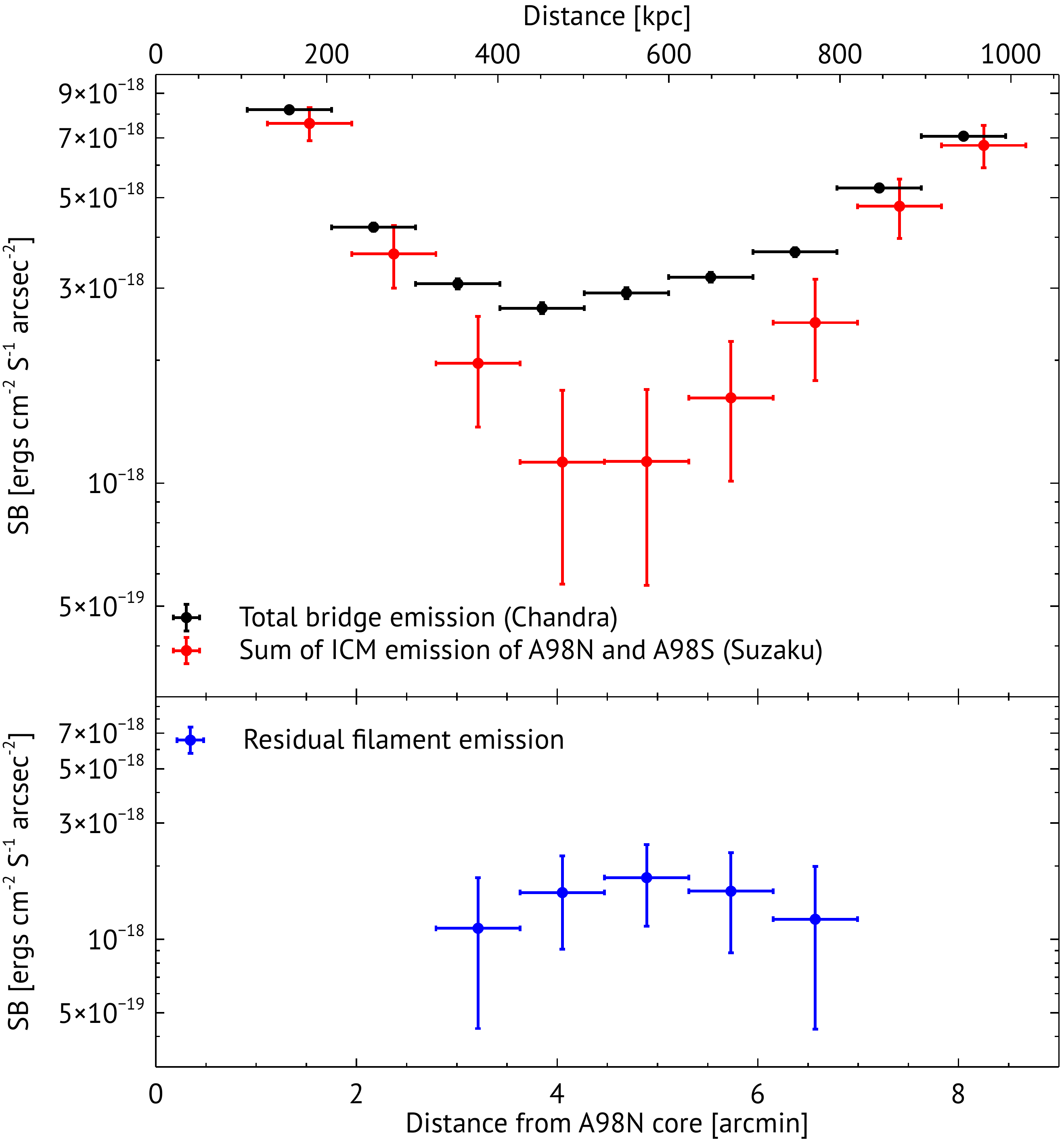}{0.52\textwidth}{(a)}
          \fig{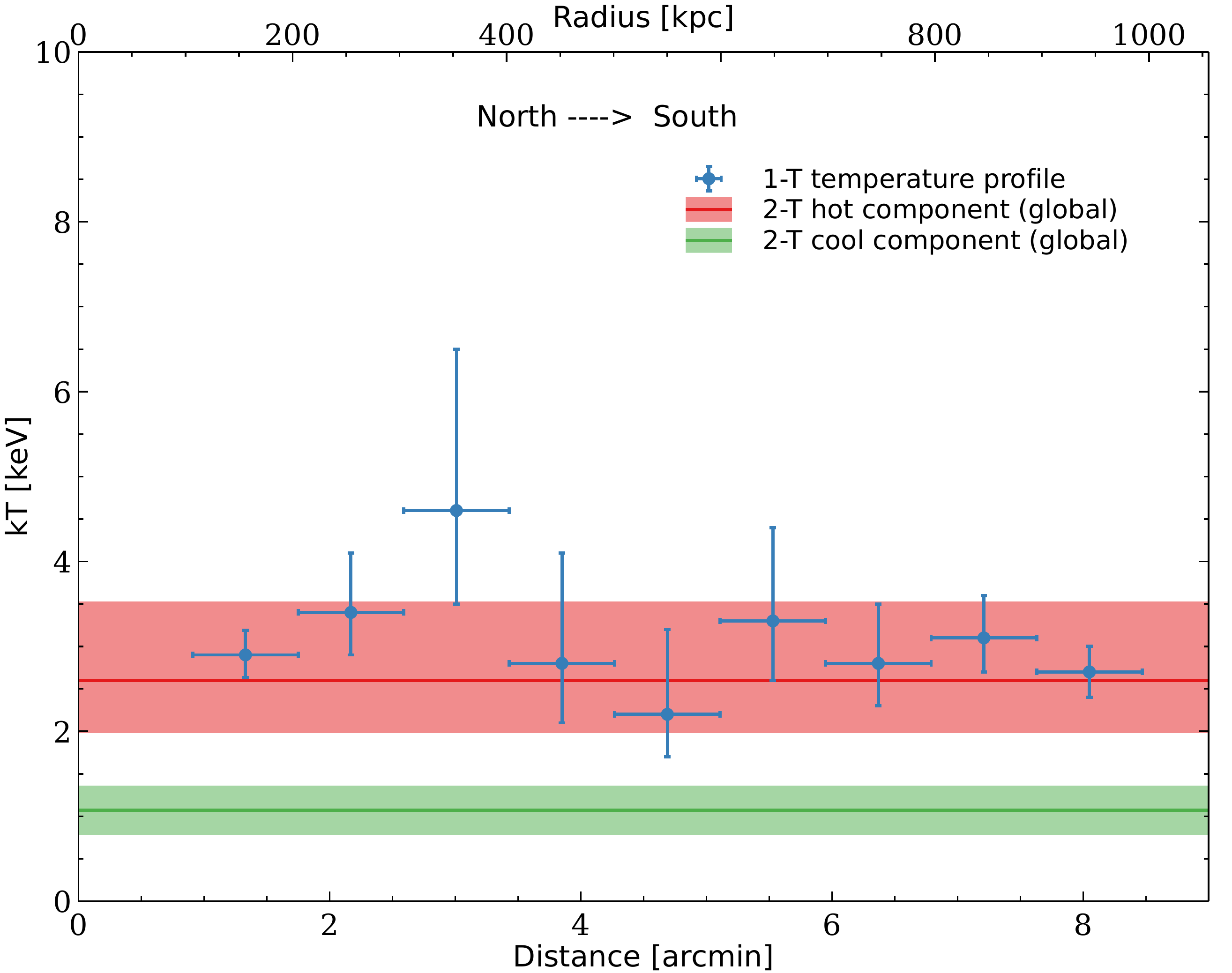}{0.52\textwidth}{(b)}
          }
\caption{
{\it left}: Black shows the 
surface brightness profile across the 
bridge measured using {\it Chandra}.
Red indicates the sum of the surface brightness profiles
of the diffuse, extended emission extracted from the
{\it Suzaku} observations.
Blue represents the residual filament emission.
{\it Right}: projected temperature profile
across the bridge, measured with a 1-T model.
Red and green shaded regions
indicate the hot
component and cool component temperatures, respectively,
measured with a 2-T model.
\label{fig:bridge_sb_temp}}
\end{figure*}

\begin{figure*}
\centering
    \includegraphics[width=0.6\textwidth]{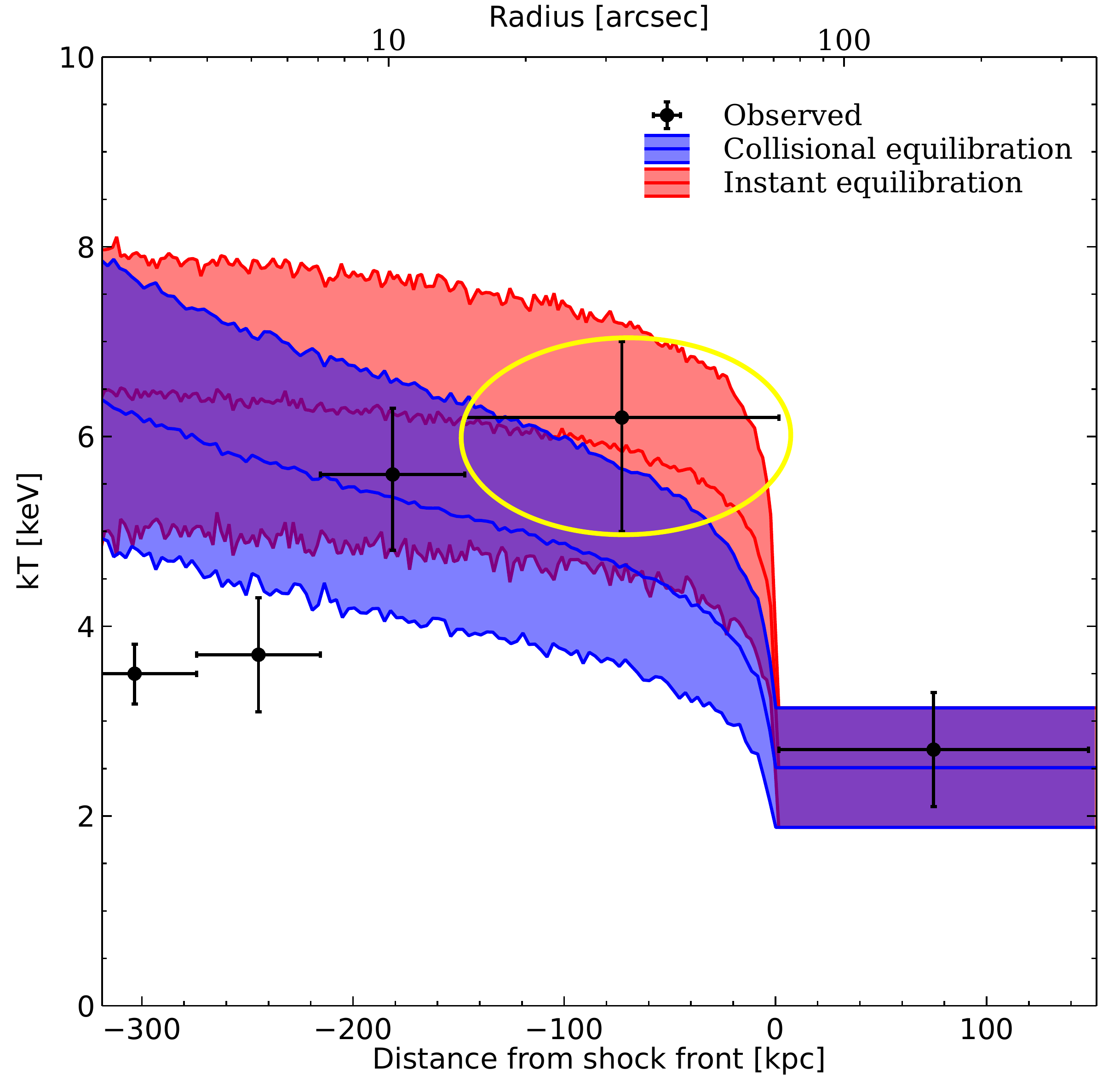}
\caption{Comparison of the temperature profile across the shock
front with the predicted electron temperature profiles based on
instant collisionless model (red) and Coulomb collisional model (blue). 
Yellow ellipse
indicates the relevant post-shock electron temperature for comparing
with two models. 
\label{fig:model}}
\end{figure*}

\section{Detection of Filament emission}
Early-stage, major mergers between 
two roughly equal mass subclusters  
are expected to typically have
their merger axis
aligned with the filaments of the 
cosmic web
(see \citet{2018ApJ...858...44A} for further discussion).
To search for faint X-ray emission 
associated with the filament
between A98N and A98S, 
we extracted a
surface brightness profile 
from nine box regions
across the bridge
between A98N and A98S
in the 0.5-2 keV energy band 
(see Figure \ref{fig:bridge}).
Figure \ref{fig:bridge_sb_temp}
shows the surface brightness profile
of the bridge. 
This surface brightness profile includes the 
emission from the extended ICM of both subclusters,
and from the filament. 
To account for only the extended ICM emission 
from both subclusters, 
we extracted
surface brightness profiles from similar regions 
placed in the opposite directions of the filament, 
using {\it Suzaku} observations of A98 
\citep{2022arXiv220608430A}, 
assuming in each that the contribution from the neighboring
subcluster is negligible.
We used {\it Suzaku} observations because
the existing 
{\it Chandra} observations do not cover 
the part of the sky needed for such analysis. 
These two diffuse surface brightness profiles are then subtracted from 
the surface brightness profile of the bridge, 
yielding the surface brightness profile of the filament.
{We use 
{\it WebPIMMS}~\footnote{\url{https://heasarc.gsfc.nasa.gov/cgi-bin/Tools/w3pimms/w3pimms.pl}}
to convert {\it Suzaku} and {\it Chandra} count rates to
physical units (erg/cm$^2$/s/arcsec$^{2}$) in the
0.5-2 keV energy band. 
We assumed a thermal {\tt APEC} model with absorption fixed to $N_{H}$=3.06$\times$10$^{20}$ cm$^{-2}$, 
abundance $Z = 0.2 Z_{\odot}$, 
and temperature $kT$ = 2.6 keV 
(as discussed later).}

Figure \ref{fig:bridge_sb_temp} shows the
resulting surface brightness profile
of the filament in the
0.5-2 keV energy band.
We detected excess X-ray surface brightness 
in the region of the bridge with 
a $\sim$3.2$\sigma$ significance level.
Similar excess emission along the bridge 
with somewhat lower significance ($\sim$ 2.2$\sigma$)
was also found by \citet{2022arXiv220608430A} using 
only {\it Suzaku} data.
Being very nearby 
to both subcluster cores, 
where emission is dominated by the
respective subcluster,
we could not detect any significant filament emission 
from the first two and last two regions.
This excess X-ray emission suggests the presence of a filament 
along the bridge connecting A98N and A98S. 

To measure the temperature across the bridge region, 
we adopted similar regions 
used for the surface brightness profile of the bridge. 
The spectra were then fitted to a single-temperature APEC model
(1-T), 
keeping the metallicity fixed to 0.2 Z$_{\odot}$. 
Figure \ref{fig:bridge_sb_temp}
shows the projected temperature profile.
The temperatures across the bridge are consistent 
within their $\sim$ 2.7$\sigma$ (90\%) uncertainty, 
except for
third region where we detected a shock. 
We next measured the global properties 
of the bridge using a 
0.6 Mpc $\times$ 0.7 Mpc box region 
(632 kpc from the A98N and 505 kpc from the A98S;
shown with cyan in Figure \ref{fig:bridge}). 
Using a single temperature emission 
model for the bridge region, we obtained a 
temperature of 1.8$^{+0.7}_{-0.4}$ keV 
and an abundance of 0.22$^{+0.12}_{-0.16}$ $Z_{\odot}$. 
Our measured temperature 
of the bridge is consistent 
with the temperatures obtained by 
\citet{2014ApJ...791..104P} using 
{\it XMM-Newton} and relatively short 
exposure {\it Chandra} observations.
We next estimated the electron
density of the bridge
using
\begin{dmath}\label{eq:ne}
    n_e = \left [1.73 \times 10^{-10} \times N\ {\rm sin}\left( {i}\right) \times (1+z)^2 \times \left(\frac{D_A}{\rm Mpc}\right)^2
    \left(\frac{r}{\rm Mpc}\right)^{-2} \left(\frac{l_{obs}}{\rm Mpc}\right)^{-1} \right]^{1/2}\ \textrm{cm}^{-3},
\end{dmath}
where $N$ is APEC normalization, 
$D_A$ is the angular diameter distance, 
$r$ and $l_{obs}$ are the radius and 
the projected length of the filament, respectively.
We obtained electron density of
$n_{e}$ = 4.2$^{+0.9}_{-0.8}$ $\times$ 10$^{-4}$ cm$^{-3}$,
assuming the filament
is in the plane of the sky ($i$ = 90$^{\circ}$) 
and has cylindrical symmetry, since for $i$ $\ll$ 90$^{\circ}$ 
we would not expect to detect a clear leading merger shock edge.
Our measured temperature and electron density are 
higher than the 
expected temperature ($\lesssim$ 1 keV) 
and electron density ($\sim$10$^{-4}$cm$^{-3}$)
for the WHIM 
\citep[e.g.,][]{2007ARA&A..45..221B,2008A&A...482L..29W,2015Natur.528..105E,2018ApJ...858...44A,2022MNRAS.510.3335H}.

The surface brightness profiles
seen in Figure \ref{fig:bridge_sb_temp} 
show that the 
emission from the filament itself 
is much fainter than the diffuse, extended cluster emission. 
A low-density gas emission is expected together with 
the ICM emission at the bridge region. 
We, therefore, adopted a two-temperature emission model
for the bridge and 
obtained a temperature of 
$kT_{hot}$ = 2.60$^{+0.93}_{-0.62}$ keV for the hotter component 
and $kT_{cool}$ = 1.07 $\pm$ 0.29 keV for the cooler component.
The higher temperature gas is
probably mostly the overlapping
extended ICM of the subclusters seen in projection.
The two-temperature model was a marginal improvement 
over the one-temperature model with an F-test probability 
of 0.08. 
The emission measure of the cooler component 
corresponds to an electron density of 
$n_{e}$ = 1.30$^{+0.28}_{-0.31}$ $\times$ 10$^{-4}$
cm$^{-3}$, 
assuming the filament is in the
plane of the sky.
From the temperature and electron density of 
the cooler component, we obtain 
an entropy of $\sim$ 416 keV cm$^2$.
Our measured temperature, density,
and entropy of the cooler component are 
consistent with what one expects for the hot, 
dense part of the WHIM 
\citep[e.g.,][]{2015Natur.528..105E,2016ApJ...818..131B,2018ApJ...858...44A,2021A&A...647A...2R}.
{\it Suzaku} observations also show
similar emission to the north of A98N, beyond the virial radius,
in the region of the putative large scale cosmic filament
\citep{2022arXiv220608430A}.
{We also check for any systematic uncertainties
in measuring the filament temperature, abundance, and density
by varying the scaling parameter of blanksky background spectra 
by $\pm$5\%. We find no significant changes.}

\section{Discussion and Conclusion}

\subsection{Nature of the shock front}
Our deep $\chandra$ observations reveal 
that the overall system is complex, 
with A98S dominated by a later-stage merger ongoing 
along the east-west direction
(Sarkar et al. in prep). 
Previously, \citet{2014ApJ...791..104P} 
argued that the surface brightness excess 
along the southern direction of A98N is more likely a 
shock with a Mach number, $\cal{M}$ $\gtrsim$ 1.3. 
With the    new data, we found that 
the temperature increases by a factor 
of $\sim$ 2.3 (from 2.7 keV to 6.4 keV) across 
the surface brightness edge, 
confirming it is a shock front
propagating along the merger axis (N/S direction).
This is the first unambiguous detection of 
axial merger shock in an early-stage merger 
{(i.e., pre-core-passage)}{,
as opposed to the late-stage 
merger
{(i.e., post-core-passage)}, 
where several previous observations
found axial shocks \citep[e.g.,][]{2010MNRAS.406.1721R,2012MNRAS.423..236R,2016ApJ...820L..20D}}.

Axial shock detection in an early-stage 
merging cluster is a 
long-standing missing piece of 
the puzzle of cluster formation. 
Previous {\it Chandra} observations of 
the pre-merger system, 1E 2216/1E 2215,
detected
an equatorial shock 
\citep{2019NatAs...3..838G}. 
Equatorial shocks 
are driven by the adiabatically expanding 
overlap region between the outskirts of the merging
subclusters. They
propagate
along the equatorial plane,
perpendicular to the merger axis.
This is in contrast to axial shocks, 
which are driven by, and ahead of, the
infalling subclusters.
\citet{2019NatAs...3..838G} did not detect any axial shock
in 1E 2216/1E 2215. 
There are conflicting findings from 
simulations on which merger shock should form first.
Recent hydrodynamical simulations 
of binary cluster mergers by \citet{2021MNRAS.506..839Z}
showed 
the formation of axial merger shocks in the 
early stages of the merging process. 
In contrast, simulations by \citet{2018ApJ...857...26H} 
indicated that equatorial shock forms long before
the axial shock.

To date, it is unclear what is driving the apparent discrepancy between
the formation of equatorial and axial shocks,
although the parameters of the merger 
(e.g., mass ratio, total mass, impact parameter) likely play a role. 
Our {\it Chandra} observation of the shock front in A98 is 
the `{first}' unambiguous 
axial shock detection in an early stage merging system, 
prior to core passage. 
We detect no equatorial shocks, but it is possible that 
they are present 
and the current observations are not
deep enough to detect them.
With this discovery, we caught two subclusters in a 
crucial epoch of the merging process, 
which will reveal any missing link to the 
shock formation process
in pre-merger systems
and provide a yardstick for future simulations.

\subsection{Electron-ion equilibrium}\label{sec:elec}
The electron heating mechanism behind a shock 
front is still up for debate. 
The collisional equilibrium model predicts that 
a shock front propagating through a collisional 
plasma heats the heavier ions dissipatively. 
Electrons are then compressed adiabatically, and subsequently 
come to thermal equilibrium with the ions via Coulomb 
collisions after a timescale defined in Equation \ref{eq:teq} 
\citep{1962pfig.book.....S,1988xrec.book.....S,1998MNRAS.293L..33E,2007PhR...443....1M}. 
By contrast, the instant equilibrium model predicts 
that electrons are strongly heated at the shock front, 
and their temperature rapidly rises to the post-shock gas temperature,
similar to the ions \citep{2006ESASP.604..723M,2012MNRAS.423..236R}. 
The electron and ion temperature jumps at the shock front are 
determined by the Rankine-Hugonoit jump conditions. 
\citet{2006ESASP.604..723M} showed that the observed temperature profile across 
the shock front of the Bullet cluster supports the instant
equilibrium model. 
\citet{2012MNRAS.423..236R} found that the temperature profile 
across one shock front of Abell~2146 
supports the collisional model while another 
shock supports the instant model.
However, in all cases the measurement 
errors prevented a conclusive determination.
{Later, \citet{2022MNRAS.514.1477R}
with deeper $\chandra$ observations
found both shocks in Abell~2146
favor the collisional model.}

Here, we compare the observed temperature profile across 
the shock front of A98 with the predicted temperature 
profiles from collisional and instant equilibrium models. 
We estimate the model electron temperatures and project 
them along the line of sight, as described in Section \ref{sec:supp_elec}. 
The resulting temperature profiles are shown in Figure \ref{fig:model}. 
The observed post-shock electron temperature
appears to be higher than the temperature 
predicted by the 
collisional equilibrium model and favors the 
instant equilibrium model, 
although we cannot rule out the collisional
equilibration model due to the large uncertainty in 
the post-shock electron temperature.

\subsection{Filament emission}
We detected 3.2$\sigma$ excess X-ray emission along the 
bridge connecting two sub-clusters (A98N and A98S), 
also detected by \citet{2022arXiv220608430A} 
(with lower significance) using $\suzaku$.
Our measured surface brightness of the cooler component ranges 
between (0.9--2.8) $\times$ 10$^{-15}$ erg cm$^{-2}$ s$^{-1}$ arcmin$^{-2}$
in the 0.5--2 keV energy band, 
which is equivalent to 
(1.2--3.6) $\times$ 10$^{-15}$ erg cm$^{-2}$ s$^{-1}$ arcmin$^{-2}$ 
in the 0.2--10 keV energy band. 
Using high-resolution 
cosmological simulations, 
\citet{2006MNRAS.370..656D} predicted that the X-ray surface 
brightness of the WHIM filaments should be 
$\sim$ 10$^{-16}$ erg cm$^{-2}$ s$^{-1}$ arcmin$^{-2}$
with a 
zero metallicity in the 0.2-10 keV energy band. 
However, an increased metallicity could induce more line emission, 
increasing the surface brightness of the filament. 
A similar conclusion was drawn by
\citet{2008A&A...482L..29W} while explaining 
their observed surface brightness of the WHIM filament, 
higher than that of cosmological simulation. 
Our measured surface brightness of the filament is 
consistent with the surface brightness 
of the WHIM filament obtained for A222/223
and A1750 \citep{2016ApJ...818..131B}. 

Using a
2--T plasma emission model,
we measured the temperature of the cooler component of
the filament, 
$kT_{\rm cool}$ = 1.07 $\pm$ 0.29 keV. 
The 2--T model was a marginal improvement 
over the 1--T model with an F-test probability 
of 0.08. 
Similar filament temperatures were 
measured for A2744 \citep[0.86--1.72 keV;][]{2015Natur.528..105E},
A222/223 \citep[0.66-1.16 keV;][]{2008A&A...482L..29W},
and A1750 \citep{2016ApJ...818..131B}.
We obtained a best-fit filament electron density of 
$n_{e}$ = 1.30$^{+0.28}_{-0.31}$ $\times$ 10$^{-4}$
cm$^{-3}$, 
assuming the filament is in the sky plane. 
If the filament has an inclination
angle ($i$) with the line of sight, the electron
density will be lower by a factor of ($\sin\ i$)$^{-1/2}$.
Previous studies also found similar electron densities for the hot, 
dense part of the WHIM in several other galaxy clusters, 
e.g., 0.88 $\times$ 10$^{-4}$ cm$^{-3}$ 
for A399/401 \citep{2021arXiv210704611H}, 
1.08 $\times$ 10$^{-4}$ cm$^{-3}$ for
A3391/3395 \citep{2018ApJ...858...44A}, 10$^{-4}$ cm$^{-3}$ for 
A2744 and A222/223 \citep{2015Natur.528..105E,2008A&A...482L..29W}.
We estimated a baryon over-density 
of $\rho/\langle \rho \rangle$ $\sim$ 240
associated with the gas in 
the filament, 
which is consistent with the expected over-density for
a WHIM filament
\citep{2007ARA&A..45..221B}. 
Assuming a cylindrical geometry for the filament, 
we estimated the associated gas mass 
to be
$M_{ gas}$ = 
3.8$^{+0.8}_{-0.6}$ $\times$ 10$^{11}$ M$_{\odot}$.

Our measured temperature and average density of the cooler component are 
remarkably consistent with what one expects for the hot, 
dense part of the WHIM, suggesting that this gas  
corresponds to the hottest, densest parts of the WHIM.
X-ray observations of the emission from WHIM filaments
are relatively rare because they have lower surface brightness
than the ICM. They offer crucial observational evidence 
of hierarchical structure formation.
Using numerical simulations, 
\citet{2001ApJ...552..473D} predicted that the gas in a WHIM filament
has a temperatures in the range 10$^{5.5}$--10$^{6.5}$ K. 
A similar conclusion was drawn by \citet{2020ApJ...889...48L} using
the kinetic S--Z effect in groups of galaxies.
Since our detected filament lies in the overlapping 
ICM of both sub-clusters, 
the gas may have been heated by the shock 
and adiabatic compression.

\begin{acknowledgments}
{We are grateful to the 
anonymous referee for 
insightful comments that greatly helped to
improve the paper.}
This work is based on observations
obtained with $\chandra$ observatory, a NASA mission
and $\suzaku$, a joint JAXA and NASA mission.
{AS and SR are
supported by the grant from NASA's  
Chandra X-ray Observatory, 
grant number GO9-20112X.}

\end{acknowledgments}

\bibliography{sample631}{}
\bibliographystyle{aasjournal}


\appendix
\section{Data reduction processes}\label{sec:supp_data}
\begin{table}[ht!]
    \centering
    \caption{{\it Chandra} observation log}
    \begin{tabular}{cccc}
\hline
Obs ID & Obs Date & Exp. time & PI \\
 & & (ks) &  \\    
\hline
11876 & 2009 Sep 17 & 19.2 & S. Murray \\
11877 & 2009 Sep 17 & 17.9 & S. Murray \\
21534 & 2018 Sep 28 & 29.5 & S. Randall \\
21535 & 2019 Feb 19 & 24.7 & S. Randall \\
21856 & 2018 Sep 26 & 30.5 & S. Randall \\
21857 & 2018 Sep 30 & 30.6 & S. Randall \\
21880 & 2018 Oct 09 & 9.9 & S. Randall \\
21893 & 2018 Nov 11 & 17.9 & S. Randall \\
21894 & 2018 Nov 14 & 17.8 & S. Randall \\
21895 & 2018 Nov 14 & 28.6 & S. Randall \\
\hline
    \end{tabular}
    \label{tab:obs_log}
\end{table}
Abell 98 was observed
twice with {\it Chandra} ACIS-I in VFAINT mode,
in September 2009 for 37 ks split into two observations 
and later in September 2018 -- February 2019 for 190 ks divided 
into eight observations. 
The combined exposure time is 
$\sim$ 227 ks (detailed observation logs are listed in Table \ref{tab:obs_log}). 
The {\it Chandra} data reduction was performed using CIAO version 
4.12 and CALDB version 4.9.4 provided by the Chandra X-ray Center (CXC). 
We have followed a standard data analysis thread
\footnote[4]{\url{http://cxc.harvard.edu/ciao/threads/index.html}}. 

All level 1 event files were reprocessed using the 
{\tt chandra$\_$repro} task, employing the latest 
gain and charge transfer inefficiency corrections,
and standard grade filtering.
VFAINT mode was applied to improve the background screening. 
The light curves were extracted and filtered using the {\tt lc$\_$clean} script to 
identify and remove periods affected by flares. The filtered exposure 
times are listed in Table \ref{tab:obs_log}. 
We used the {\tt reproject$\_$obs} task to reproject 
all observations to a common tangent position and combine them. 
The exposure maps in the 0.5-2.0 keV energy bands were created using 
the {\tt flux$\_$obs} script by providing a weight spectrum. 
The weight spectrum was generated using the {\tt make$\_$instmap$\_$weights} 
task with an absorbed APEC plasma emission model and  
a plasma temperature of 3 keV. 
To remove low signal-to-noise areas near chip edges and chip gaps,
we set the pixel value to zero for those pixels with an exposure of 
less than 15\% of the combined exposure time. 

Point sources were identified using {\tt wavdetect} with a range of
wavelet radii between  1--16 pixels to maximize the 
number of detected point sources. We set the detection 
threshold to $\sim$ 10$^{-6}$, for which we expect
$\lesssim$ 1 spurious source 
detection per CCD. 
We used blank-sky observations provided by the CXC 
to model the non-X-ray 
background and emission from 
the foreground structures 
(e.g., Galactic Halo and Local Hot Bubble) along the observed direction. 
The blanksky--background was generated using the {\tt blanksky} task and then 
reprojected to match the coordinates of the observations. 
The resulting blanksky-background was 
normalized so that the hard band 
(9.5 - 12 keV) count rate matched the observations.

\section{Electron heating mechanism}\label{sec:supp_elec}
{When a shock front propagates through a plasma, 
it heats the ions dissipatively in a layer 
that has a width of few 
ion-ion collisional mean free paths. 
On the other hand, 
having very high thermal velocity compared to the shock, 
the electron temperature does not 
jump by the same high factor as the ion temperature.
The electrons are compressed adiabatically 
at first in merger shocks and subsequently 
equilibrate with the ions via Coulomb scattering 
after a timescale \citep{1962pfig.book.....S,1988xrec.book.....S} 
given by,
\begin{equation} \label{eq:teq}
    {t_{eq}}( e,p) \approx 6.2 \times 10^8 {\rm yr} \left( \frac{{T}_{ e}}{10^8 \rm K} \right)^{3/2} \left( \frac{{ n}_{ e}}{10^{-3} {\rm cm}^{-3}} \right)^{-1}
\end{equation}
where $T_{e}$ and $n_{e}$ are the
electron temperature and density, respectively.

Alternatively, the instant 
collisionless shock model
predicts that 
the electrons and ions reach thermal equilibrium 
on a timescale much shorter than $t_{ eq}$ after passing the shock front, 
where the post-shock electron temperature is determined 
by the pre-shock electron temperature and the RH jump conditions
\citep{2007PhR...443....1M}. 
We use the best-fit density profile shown in 
Figure \ref{fig:image_edge_fitting} to project this model electron temperature 
along the line of sight analytically \citep{2012MNRAS.423..236R}. 

In the collisional equilibration model, 
the electron temperature rises at the shock front 
through adiabatic compression,
\begin{equation}\label{eq:Te} 
    { T}_{ e,2} = { T}_{ e,1} \left( \frac{\rho_2}{\rho_1} \right)^{\gamma -1}
\end{equation}
where $\rho_{1}$ and $\rho_{2}$ are the gas density
in the pre-shock and post-shock regions, and 
$\gamma$ = 5/3. Electron and ion temperatures then 
subsequently equilibrate via Coulomb collision at a rate
given by, 
\begin{equation}\label{eq:rate}
    \frac{ dT_e}{ dt} = \frac{{ T_i} - { T_e}}{ t_{eq}} 
\end{equation}
where $T_{ i}$ is the ion temperature.
Since the total kinetic energy density is conserved, 
the local mean gas temperature, $T_{ gas}$ is constant with time, 
where $T_{ gas}$ is given by,
\begin{equation}\label{eq:Tgas}
    { T_{gas}} = \frac{ n_eT_e +n_iT_i}{ n_i + n_e} = \frac{ 1.1T_e + T_i}{2.1}
\end{equation}
where $n_{i}$ is the ion density.

We integrate Equation \ref{eq:rate}
by using Equation \ref{eq:teq} and \ref{eq:Tgas}
to obtain the model electron temperature analytically.
Finally, we project the model electron temperature profile 
along the line of sight \citep{1998MNRAS.293L..33E} by,
\begin{equation} \label{eq:project}
    \langle { T} \rangle =\int_{b^2}^{\infty}\frac{\epsilon( r){ T_e( r) dr^2}}{\sqrt{r^2 - b^2}} \bigg/ {\int^{\infty}_{b^2}\frac{\epsilon( r){ dr^2}}{\sqrt{r^2 - b^2}}}
\end{equation}
where $\epsilon(r)$ 
is the emissivity at physical radius $r$
and $b$ is the distance from the shock front.

The emissivity-weighted electron temperature 
should be close to what one observes 
with a perfect instrument with a flat energy 
response across the relevant energy range.
Since this is not the case, we
convolve the instant and collisional model 
electron temperatures
with the response of the telescope to predict 
what we expect to measure \citep{2012MNRAS.423..236R}. 
We first estimate the emissivity-weighted electron temperature
using the above models and corresponding emission 
measure using the best-fit density discontinuity model
(Equation \ref{eq:ne})
in a small volume dV. 
We then sum the emission measures with similar temperatures 
for each annulus using a temperature binning of 0.1 keV. 
We simulate spectra in XSPEC using a multi-temperature 
absorbed {\tt apec} model. 
We fix the temperature of each component with the median 
temperature of each bin and 
calculate the normalization based on 
the summed emission measure in that temperature bin.
We also fix the abundance to 0.4 $Z_{\odot}$, n$_{\rm H}$ to galactic value, 
and redshift to 0.1043. 
To estimate the expected projected temperature 
in this annulus,
we finally fit this simulated spectra with a single temperature 
absorbed {\tt apec} model with abundance, n$_{\rm H}$, and 
redshift fixed as above.

We adopt the Monte Carlo technique with 1000 realizations 
to measure uncertainty in model electron temperatures. 
Assuming a Gaussian distribution of the pre-shock temperature, 
which is the largest source of uncertainty, 
we repeated the model and projected temperature 
calculations 1000 times with a new value 
of pre-shock temperature each time. 
The resulting instant and collisional model 
temperature profiles with 1$\sigma$ uncertainty
are shown in Figure \ref{fig:model}.}




\end{document}